\documentclass[prl,preprintnumbers,amsmath,amssymb,superscriptaddress,groupedaddress,nofootinbib,tightenlines,twocolumn]{revtex4}

\usepackage{graphicx}
\usepackage{sidecap}
\usepackage{amsmath}
\usepackage{bm}
\usepackage{bbm}

\def\beqn{\begin{eqnarray}}
\def\eeqn{\end{eqnarray}}
\def\barr{\begin{array}}
\def\earr{\end{array}}
\def\btab{\begin{tabular}}
\def\etab{\end{tabular}}
\def\bite{\begin{itemize}}
\def\eite{\end{itemize}}
\def\bcen{\begin{center}}
\def\ecen{\end{center}}

\def\eq{\begin{equation}}
\def\ee{\end{equation}}
\def\nn{\nonumber}

\def\pdagger{p\hspace{-0.17cm}/}

\def\qdagger{q\hspace{-0.18cm}/}
\def\keldagger{k\hspace{-0.2cm}/}

\def\q2dagger{q_2\hspace{-0.35cm}/\;}

\newcommand{\be}{\begin{equation}}
\newcommand{\bea}{\begin{eqnarray}}
\newcommand{\eea}{\end{eqnarray}}

\begin{document}
\title{$\gamma W$-box 
Inside-Out: 
Nuclear Polarizabilities Distort the Beta Decay Spectrum}
\author{Mikhail Gorchtein} 
\email{gorshtey@uni-mainz.de}
\affiliation{Helmholtz-Institut Mainz, Mainz, Germany\\ 
PRISMA Cluster of Excellence, Johannes Gutenberg-Universit\"at, Mainz, Germany}

\begin{abstract}
I consider the $\gamma W$-box correction to superallowed nuclear $\beta$-decays in the framework of dispersion relations. I address a novel effect of a distortion of the emitted electron energy spectrum by nuclear polarizabilities and show that this effect, while neglected in the literature, is sizable. 
The respective correction to the $\beta^+$ spectrum is estimated to be $\Delta_R(E)=(1.6\pm1.6)\times10^{-4}{E}/{\rm MeV}$ assuming a conservative 100\% uncertainty.  The effect is positive-definite and can be observed if a high-precision measurement of the positron spectrum is viable. If only the full rate is observed, it should be included in the calculated ${\cal F}t$-values of nuclear decays. I argue that this novel effect should be 
included in the analyses of nuclear beta decay experiments to ensure the correct extraction of $V_{ud}$ from decay rates, and of the Fierz interference term from precision measurements of decay spectra.
\end{abstract}
\date{\today}
\maketitle

The observation of $\beta$-decay has furnished the evidence for many fundamental ingredients of the Standard Model (SM). Universality of the weak interaction and conservation of the vector current (CVC) led to the introduction of the Cabibbo-Kobayashi-Maskawa (CKM) quark mixing matrix which has to obey the constraint of unitarity. 
Unitarity of its first row, $|V_{ud}|^2+|V_{us}|^2+|V_{ub}|^2=0.9994(5)$ \cite{PDG2018} is one of the most stringent constraints on the parameters of SM and its extensions~\cite{Gonzalez-Alonso:2018omy}. 

The top left corner element $|V_{ud}|=0.97420(21)$ \cite{PDG2018} dominates both the value and the uncertainty of the unitarity constraint, and is obtained almost exclusively from the global analysis of a number of superallowed $0^+-0^+$ $\beta$-decays \cite{Hardy:2014qxa,Hardy:2018zsb}. One of the cornerstones of this analysis is an adequate calculation of one-loop radiative corrections 
which have been studied for over 6 decades, and the formalism has been worked out, e.g., in Refs. \cite{Sirlin:1967zza,Marciano:2005ec}. 

The very accurate extraction of $V_{ud}$ from superallowed nuclear decays is empowered by the following formula,
\begin{eqnarray}
\left|V_{ud}\right|^2=2984.43s/\big[\overline{{\mathcal{F}t}}(1+\Delta_R^V)\big]. \label{eq:1}
\end{eqnarray}
The radiative correction $\Delta_R^V$ is evaluated on a free neutron \cite{Marciano:2005ec}, and is conventionally singled out also for nuclear decays. 
The universal and very precise value $\overline{\mathcal{F}t}=3072.07(63)$s is an average of 14 reduced half-lives \cite{Hardy:2014qxa,Hardy:2018zsb}
\beqn
\mathcal{F}t=ft(1+\delta'_R)(1+\delta_{NS}-\delta_C),\label{eq:2}
\eeqn
which are obtained from the measured half-lives $t$ and calculated statistical factors $f$, and should be independent of the particular decay as a consequence of CVC. The ``outer" correction  $\delta'_R$ depends on the emitted electron energy and the charge of the daughter nucleus. I refer the reader to Ref. \cite{Hayen:2017pwg} for a recent review of energy-dependent corrections. The nuclear structure dependence resides in the energy-independent ``inner" corrections $\delta_C$ and $\delta_{NS}$: the former stems from isospin-breaking corrections to the tree-level matrix element of the Fermi operator, and the latter from nuclear effects in the $\gamma W$-box, defined with respect to the free-neutron $\gamma W$-box entering $\Delta_R^V$.

The $\gamma W$-box plays a central role in the uncertainty of $V_{ud}$. Recently, it was re-examined in the dispersion relation (DR) framework \cite{Seng:2018yzq,Seng:2018b}. Ref. \cite{Seng:2018yzq} addressed hadronic contributions to the universal correction $\Delta_R^V$, and found a substantial shift in the extracted value of $V_{ud}$ with a reduced hadronic uncertainty, $|V_{ud}|=0.97370(14)$, raising tension with unitarity,  
$|V_{ud}|^2+|V_{us}|^2+|V_{ub}|^2=0.9984(4)$.
Consequently, Ref. \cite{Seng:2018b} investigated the robustness of the procedure of splitting the $\gamma W$-box on a nucleus into the universal, free-neutron $\Delta_R^V$ and the nucleus-specific $\delta_{NS}$. 
In particular, the ``quenching" of the free-nucleon elastic box contribution was addressed,
and a dispersive evaluation 
suggested that this effect previously calculated in Ref. \cite{Towner:1994mw} and included in all subsequent analyses of the superallowed nuclear decays was underestimated. A proper account of the quasielastic contribution led to a reduction in the reduced half-life, $\overline{{\cal F}t}=3072.07(63)\,$s $\to\,\overline{{\cal F}t}^{\rm new}=3070.5(1.2)\,$s, bringing $V_{ud}$ closer to its old value, $|V_{ud}|=0.97395(21)$ and improving the agreement with unitarity somewhat, $|V_{ud}|^2+|V_{us}|^2+|V_{ub}|^2=0.9989(5)$.

This Letter is dedicated to a critical assessment of yet another ingredient of Eqs. (\ref{eq:1},\,\ref{eq:2}), the splitting of the full radiative correction into ``inner" and ``outer". The logics behind this splitting uses the fact that while the energy released in superallowed decays is few MeV, the scale that governs the strong interaction is the pion mass $M_\pi\approx140$ MeV. Then, energy-dependent effects due to strong interaction will only show up at ${\alpha\over\pi} {E\over M_\pi}\sim10^{-5}$ (with $\alpha\approx1/137$ the fine structure constant and $E$ the energy of the lepton), negligible at the present level of precision. 

All references that have dealt with nuclear structure contributions to the $\gamma W$-box in the past have assumed the correctness of this argument. However, the presence of the nuclear excitation spectrum that is separated from the ground state by only a few MeV provides a more natural energy scale $\Lambda_{\rm Nucl.}$, generically expected to lie between the two extremes, $Q < \Lambda_{\rm Nucl.}< M_\pi$. It is then possible that energy- and nuclear structure-dependent corrections scale as ${\alpha\over\pi} {E\over \Lambda_{\rm Nucl.}}\sim 10^{-3}-10^{-5}$, depending on the exact value of that scale. If it is small enough, 
the conventional splitting of the $\gamma W$-box into inner and outer contributions is not warranted (the inner correction leaks into the outer one), and along with the pure QED outer correction $\delta'_R$, a new energy-dependent nuclear structure correction $\delta_E^{NS}$ has to be included in the universal ${\cal F}t$ value.
I investigate this scenario below.

\begin{figure}[h]
\begin{center}
\includegraphics[width=7.0cm]{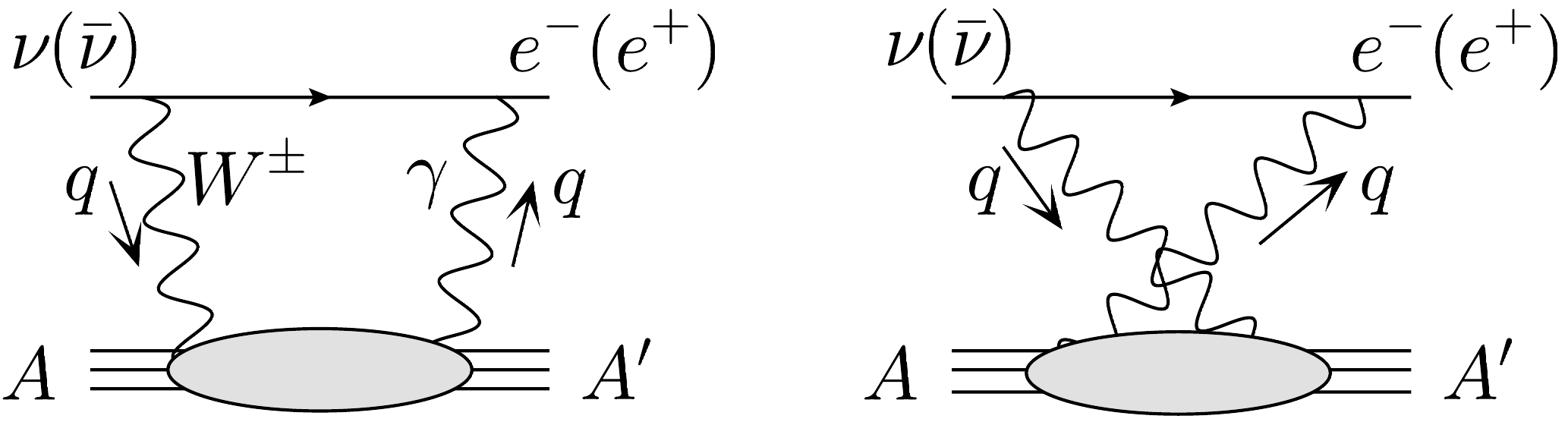}
\caption{The direct and crossed $\gamma W$-box diagrams for a nuclear $\beta^-$ ($\beta^+$) decay of the parent $A$ into the daughter nucleus $A'$ with the emission of an electron (positron), respectively.}
\label{fig:box}
\end{center}
\end{figure}
I consider a forward scattering process $\{\nu/\bar\nu\}(k)+A(p)\to e^\mp(k)+A'(p)$ in the limit of zero momentum transfer (no nuclear recoil) but finite lepton and neutrino energies. At this stage, the masses of the initial and final nuclear states $A,A'$ are taken equal, $M$, and electron and neutrino masses are neglected. 
This is an adequate approximation for the purposes of this study. 
The $\gamma W$-box amplitude in Fig.~\ref{fig:box} is defined as 
\beqn
T_{\gamma W}=\frac{\alpha G_FV_{ud}}{\sqrt2}\!\!
\int\!\!\frac{d^4q}{\pi^2}\frac{\bar u_e\gamma_\nu(\keldagger-\qdagger)\gamma_\mu(1-\gamma_5)u_\nu}{q^2(k-q)^2(-1+q^2/M_W^2)}
W_{\gamma W}^{\mu\nu},\label{eq:gaWbox}
\eeqn
with $q$ the 4-momentum of the $\gamma$ and $W^\pm$ boson.
The spin-independent Compton tensor takes the form,
\beqn
W^{\mu\nu}_{\gamma W}=
-g^{\mu\nu}
\eta T_1
+\frac{p^\mu p^\nu}{(p\cdot q)}
\eta T_2
+\frac{i\epsilon^{\mu\nu\alpha\beta}p_\alpha q_\beta}{2(p\cdot q)}T_3,\label{eq:Wmunu}
\eeqn
with the phase $\eta=\pm1$ for the $\beta^\pm$ process, respectively. 
The forward amplitudes are functions of the photon energy $\nu={(p\cdot q)\over M}$ and photon virtuality $Q^2$, 
and are related to the inclusive structure functions via Im$\,T_i=F_i$ defined via the commutator of the hadronic electromagnetic and charged weak currents $J^\nu_{em}$ and $J^{\pm,\mu}_W$, respectively,
\beqn
&&\frac{i}{4\pi}\int dxe^{iqx}\langle A'|[J^\nu_{em}(x),J^{\pm,\mu}_W(0)]|A\rangle\nn\\
&&=-g^{\mu\nu}
F_1
+\frac{p^\mu p^\nu}{(p\cdot q)}
F_2
+\frac{i\epsilon^{\mu\nu\alpha\beta}p_\alpha q_\beta}{2(p\cdot q)}T_3.
\eeqn
I define the $\Box_{\gamma W}$-correction per active nucleon as
\beqn
T_{W}+T_{\gamma W}=-N{\sqrt2}{G_FV_{ud}}\left[1+\Box_{\gamma W} \right]\bar u_e\pdagger (1-\gamma_5)u_{\nu},
\label{eq:boxdef}
\eeqn
with $N$ the number of active nucleons. 

The imaginary part of the box diagram is easily obtained using definitions in Eqs. (\ref{eq:gaWbox}-\ref{eq:boxdef}), 
\beqn
&&{\rm Im}\,\Box_{\gamma W}(E)=\frac{\alpha}{N}\int_{0}^{E_1^{m}}\frac{dE_1}{E}\int_{0}^{Q_{m}^2}dQ^2
\\
&&\!\!\!\!\!\!\times\left[\frac{F_3}{2M\nu}\left(1-\frac{\nu}{2E}\right)+\frac{\eta F_1}{2ME}+\left(\frac{E-\nu}{\nu Q^2}-\frac{1}{4E\nu}\right)\eta F_2\right].\nn
\eeqn
Energy variables appearing above are defined in terms of the invariants $s=(p+k)^2$, $W^2=(p+q)^2$  as $E={s-M^2\over 2M}$, $E_1={s-W^2\over 2M}$, $\nu={W^2-M^2+Q^2\over 2M}$. 
The upper limits are $E_1^{m}=E-\epsilon$, with $\epsilon$ the threshold for the photobreakup of the target nucleus, and $Q_{m}^2=(s-W^2)(s-M^2)/s$. The real part of the box correction is obtained from the forward dispersion relation of the form
\beqn
{\rm Re}\,\!\Box_{\gamma W}(E)=\frac{1}{\pi}\int\limits_{\epsilon}^\infty
\left[\frac{dE'}{E'-E}\pm\frac{dE'}{E'+E}\right]
{\rm Im}\,\!\Box_{\gamma W}(E'),
\eeqn
with the first and second terms in the square bracket originating from the discontinuity of the direct and crossed graph, respectively. 
The sign between the two depends on the isospin structure of the Compton amplitudes. Electromagnetic interaction does not conserve isospin, and $T_i$ contain two isospin components, 
$T_i^{(0)}\tau^a+T_i^{(-)}\frac{1}{2}[\tau^3,\tau^a]$ 
which behave differently under crossing, 
\beqn
T_i^{(0,-)}(-\nu,Q^2)=\xi_i^{(0,-)}T_i^{(0,-)}(\nu,Q^2),
\eeqn
with 
$\xi^{(0)}_i=1$ for $T_1$ and $\xi^{(0)}_i=-1$ for $T_{2,3}$, and $\xi^{(-)}_i=-\xi^{(0)}_i$. As a result, the $\gamma W$-box will contain both even and odd powers of energy.
I account for the leading $E$-dependence: constant in the $E$-even and linear in the $E$-odd pieces, respectively. 
The hadronic structure-dependent part of the $E$-even piece that is due to the weak vector current (contribution of $F_{1,2}^{(-)}$) cancels against other 1-loop corrections \cite{Sirlin:1967zza} and is omitted. To reflect this subtraction I use notation $\overline{\Box_{\gamma W}^{even}}$.
Changing the order of integration and assuming that the energy released in the $\beta$-decay process is smaller than nuclear excitations, I obtain the dispersion representation for the leading $E$-behavior of the $\gamma W$-box:
\beqn
{\rm Re}\,\overline{\Box_{\gamma W}^{even}}&=&\frac{\alpha}{\pi N}\int\limits_{0}^{\infty}dQ^2\int\limits_{\nu_{thr}}^\infty{d\nu}\nn
\frac{F_3^{(0)}}{M\nu}\left(\frac{1}{E_{min}}-\frac{\nu}{4E_{min}^2}\right),\nn
\eeqn
\beqn
&&{\rm Re}\,\Box_{\gamma W}^{odd}=\frac{\alpha E}{3\pi NM}\int\limits_{0}^{\infty}dQ^2\int\limits_{\nu_{thr}}^\infty\frac{d\nu}{E_{min}^3}\Big[{\eta F_1^{(0)}}\label{eq:Boxodd}\\
&&
+\left.\frac{M}{\nu}
\left(\frac{3\nu E_{min}}{Q^2}+1
\right)\eta F_2^{(0)}+\frac{\nu+3\sqrt{\nu^2+Q^2}}{4\nu}F_3^{(-)}\right],\nn
\eeqn
where 
$E_{min}={\nu+\sqrt{\nu^2+Q^2}/2}$ 
and $\nu_{thr}=\epsilon+{Q^2/2M}$. The $E$-even piece has recently been addressed in \cite{Seng:2018yzq,Seng:2018b}, and will be discussed in the text around Eq. (\ref{eq:e-indep}). In the rest of the letter I concentrate on the $E$-odd part and estimate its size in two different models.\\

\noindent
{\it Dimensional analysis with the photonuclear sum rule}\\
The photonuclear sum rule expresses the electric dipole polarizability $\alpha_E$ as an integral over electromagnetic structure functions $F_{1,2}$
\beqn
\alpha_E=\frac{2\alpha}{M}\int\limits_\epsilon^\infty\frac{d\nu}{\nu^3}F_1(\nu,0)
=2\alpha\int\limits_\epsilon^\infty\frac{d\nu}{\nu^2}\frac{\partial}{\partial Q^2}F_2(\nu,0).
\eeqn
The equality between the representations with $F_1$ and the $Q^2$-slope of $F_2$ is a reflection of gauge invariance. The electromagnetic structure functions should be similar to their vector charged current - electromagnetic current interference counterpart. 
I next assume the very low $Q^2$ under the integral to dominate (hence $E_{\min}\to\nu$), and the $Q^2$ dependence of the dipole polarizability to follow that of the nuclear form factor $\sim e^{-R_{Ch}^2Q^2/6}$. Discarding the contribution of $F_3$ for which no information in terms of  polarizabilities is available, I obtain for the $\beta^+$ case
\beqn
{\rm Re}\,\Box_{\gamma W}^{odd}\sim ({4\alpha_E}/{\pi NR_{Ch}^2})E
\eeqn
The observed approximate scaling of nuclear radii with the atomic number $R_{Ch}\sim R_0 A^{1/3}$ with $R_0\approx1.2$ fm \cite{DeJager:1974liz}, and that of the nuclear electric dipole response $\alpha_E\sim(2.2\times10^{-3})A^{5/3}$ fm$^3$ 
\cite{Berman:1975tt}, leads to 
an $E$-dependent correction to the differential decay rate, 
\beqn
\delta_{NS}(E)=2{\rm Re}\,\Box_{\gamma W}^{odd}(E)=2\times10^{-5}\left({E\over{\rm MeV}}\right){A\over N},\label{eq:est1}
\eeqn
Note that for all measured superallowed decays $A/N\approx2$. \\

\noindent
{\it Estimate in the free Fermi gas model}\\
In a microscopic picture, a large part of the nuclear polarizability can be explained by the quasielastic mechanism. 
\begin{figure}[h]
\includegraphics[width=3.5cm]{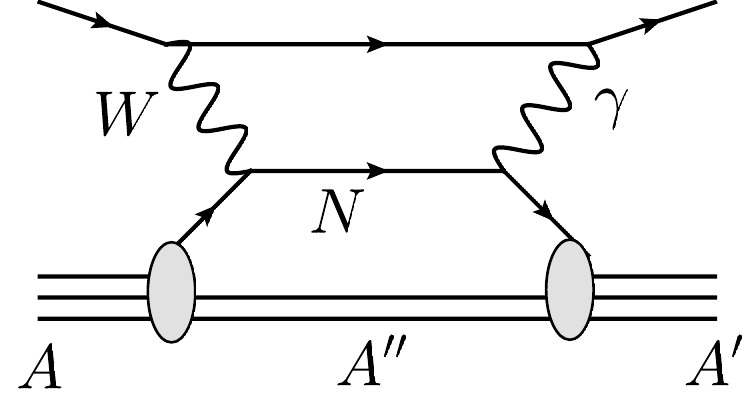}
\caption{Quasielastic contribution to the nuclear $\gamma W$-box.}
\label{fig:QE}
\end{figure}
The generalized Compton reaction on a nucleus proceeds via the knockout of a single active nucleon by the initial electroweak probe, leaving the remaining part of the nucleus unaffected, and the reabsorption of the nucleon back into the nucleus accompanied by the emission of the final photon, see Fig. \ref{fig:QE}. 
The finite gap between the bound state and the continuum (removal energy) and the Fermi momentum $k_F$, the typical momentum of a nucleon inside the nucleus, are the two relevant parameters that govern the size of the nuclear polarizability. 
In the case of a decay process, the initial and final states are not identical due to the $n\to p(p\to n)$ conversion for the $\beta^-$($\beta^+$) process, respectively. Apart from the change of the nucleon specie and the charge of the nucleus in the initial (parent) and final (daughter) state, the mass of the daughter is smaller, which is a prerequisite for the decay to take place. For the quasielastic process $W^\pm+A\to n(p)+A''\to\gamma+A'$, with $A''$ a spectator nuclear state, there are two distinct removal energies at the first and the second stage of the reaction. Specifically for the $\beta^+$ process, $\epsilon_1=M_{A''}+M_n-M_A$ and $\epsilon_2=M_{A''}+M_n-M_{A'}$ obeying $\epsilon_2>\epsilon_1$. In the recent work \cite{Seng:2018b} it was proposed to use an effective removal energy $\bar\epsilon=\sqrt{\epsilon_1\epsilon_2}$ to simplify the calculation. For the 20 superallowed $\beta^+$ decays listed in \cite{Hardy:2014qxa} the effective removal energies fall within a narrow range, $\bar\epsilon=7.5\pm1.5$ MeV \cite{Seng:2018b}. In the free Fermi gas (FFG) model the structure functions entering Re$\,\Box_{\gamma W}$ have a generic form
\beqn
({1}/{N})F_i(\nu,Q^2)=f_i^B(Q^2)S(\nu,Q^2,\bar\epsilon,k_F),
\eeqn
with the spectral function 
\beqn
S=F_{\mathrm P}(|\vec q|,k_F) \int d^3\vec k\,|\phi(k)|^2\delta((k+q)^2-M^2).\label{eq:QE}
\eeqn
Above, $k$ is the 4-momentum of the active nucleon, $\phi(k)$ the momentum distribution in the FFG model, $|\phi(k)|^2=3/ (4\pi k_F^3)\theta(k_F-k)$ normalized as $ \int d^3\vec k|\phi(k)|^2=1$. Pauli blocking is described by the Pauli function 
\beqn
F_{\mathrm P}(|\vec q\,|,k_F)=\frac{3|\vec q\,|}{4k_F}[1-\vec q\,^2/(12 k_F^2)]\;\; {\rm for}\;|\vec q|\leq 2k_F,
\eeqn
and $F_{\mathrm P}=1$ otherwise, and $|\vec q\,|=\sqrt{\nu^2+Q^2}$ stands for the 3-momentum of the virtual photon ($W^\pm$ boson). The $\delta$-function reflects the knock-out nucleon being on shell. The integral in Eq. (\ref{eq:QE}) can be carried out analytically \cite{Seng:2018b} after which the dependence of the spectral function $S$ on the breakup threshold becomes explicit. 
Finally, the residues $f_i$ corresponding to the coefficient in front of the $\delta$ function in the nucleon Born contribution read
$f_1^{(0)}=(Q^2/8)G_M^WG_M^S$,  $f_2^{(0)}=({Q^2}/{4})[{G_E^VG_E^S+\tau G_M^VG_M^S}]/({1+\tau})$, and 
$f_3^{(-)}=-({Q^2/4})G_AG_M^V$,
with $G_{E,M}^{S,V}=G_{E,M}^p(Q^2)\pm G_{E,M}^n(Q^2)$ the nucleon isoscalar and isovector electromagnetic form factors, the axial form factor $G_A$ with $G_A(0)=-1.2755$, and the nucleon recoil $\tau=Q^2/4M_p^2$. A numerical evaluation with the effective separation energy $\bar\epsilon=7.5\pm1.5$ MeV and Pauli momentum $k_F=235\pm10$ MeV leads to 
\beqn
\delta_{NS}(E)=(2.8\pm0.4)\times10^{-4}\left({E}/{{\rm MeV}}\right).
\eeqn
This estimate is one order of magnitude larger than the naive estimate with the nuclear electric dipole polarizability and the nuclear size. It is well known that QE cross sections with slightly virtual photons are much larger than with real photons, so the estimate $\alpha_E(Q^2)\sim\alpha_E(0)e^{-R_{Ch}^2Q^2/6}$ used in the previous section is likely to underestimate the actual effect. On the other hand, the FFG model is known to overestimate the quasielastic response at very low values of $Q^2$ where meson exchange currents tend to lead to a suppression. So the realistic size of the effect is likely to lie between those two extremes. Note that the contribution of $F_3^{(-)}$ dominates over the other two terms in Eq. (\ref{eq:Boxodd}) in FFG due to the large nucleon isovector magnetic moment.\\

%
\noindent
{\it Numerical results and the effect on the ${\cal{F}}t$-values}\\
Above, I obtained the energy-dependent correction in two different models which give an estimate of the lower and upper bound of the effect. For numerical analysis I will use their average with a 100\% uncertainty,
\beqn
\delta_{NS}(E)\sim(1.6\pm1.6)\times10^{-4}\left({E}/{{\rm MeV}}\right).\label{eq:DRE}
\eeqn
This result is independent of the nucleus and directly observable if the $\beta$-spectrum is measured. If only the total rate is observed, the respective correction  to the ${\cal F}t$-value is obtained by integrating $\delta_{NS}(E)$ over the $\beta$-spectrum. This correction will be decay-specific via the $Q$-value, $Q=M_A-M_{A'}$, with $M_A\,(M_{A'})$ the mass of the parent (daugther) nucleus, respectively. It is defined as 
\beqn
\delta^E_{NS}=\int\limits_{m_e}^{E_m}dE \rho_0(E,E_m)\delta_{NS}(E)/\int\limits_{m_e}^{E_m}dE \rho_0(E,E_m),
\eeqn
with $\rho_0=Ep(E_m-E)^2$ the tree-level decay spectrum function, $p=\sqrt{E^2-m_e^2}$ the electron 3-momentum, $m_e$ the electron mass, and $E_m=(M_A^2-M_{A'}^2+m_e^2)/2M_A\approx Q$ the maximal electron energy available in a given decay. The integration with the estimate of Eq. (\ref{eq:DRE}) leads to 
\beqn
\delta^E_{NS} \approx (8\pm8)\times10^{-5} ({Q}/{{\rm MeV}}),
\eeqn
which modifies the ${\cal F}t$ values as 
${\cal F}t' = {\cal F}t(1+\delta^E_{NS})$.
The shift due to the nuclear polarizability contribution $\delta{\cal F}t={\cal F}t\times\delta_E^{NS}$ 
is shown for the 14 most accurately measured superallowed decays in Table \ref{tab1} along with the central values and the respective uncertainties of the original analysis of Ref. \cite{Hardy:2014qxa}. 
\begin{table}[h]
  \begin{tabular}{l|c|c|c|r}
\hline
Decay & $Q$ (MeV) &  $\delta^E_{NS}(10^{-4})$ & $\delta {\cal F}t$(s) & ${\cal F}t$(s) \cite{Hardy:2014qxa} \\
\hline
\hline
$^{10}C$
&  1.91  &  1.5 & 0.5 & 3078.0(4.5)\\
$^{14}O$
&  2.83  &  2.3 & 0.7 & 3071.4(3.2) \\
$^{22}Mg$
&  4.12  &  3.3 & 1.0 & 3077.9(7.3)\\
$^{34}Ar$
&  6.06  &  4.8 & 1.5 & 3065.6(8.4) \\
$^{38}Ca$
&  6.61  &  5.3 & 1.6 & 3076.4(7.2) \\
$^{26m}Al$
&  4.23  &  3.4 & 1.0 & 3072.9(1.0) \\
$^{34}Cl$
&  5.49  &  4.4 & 1.4 & 3070.7$^{+1.7}_{-1.8}$\\
$^{38m}K$
&  6.04  &  4.8 & 1.5 & 3071.6(2.0) \\
$^{42}Sc$
&  6.43  &  5.1  & 1.6 & 3072.4(2.3) \\
$^{46}V$
&  7.05  &  5.6  & 1.7 & 3074.1(2.0) \\
$^{50}Mn$
&  7.63  & 6.1 &  1.9 & 3071.2(2.1) \\
$^{54}Co$
&  8.24  &  6.6  & 2.0 & 3069.8$^{+2.4}_{-2.6}$\\
$^{62}Ga$
&  9.18  &  7.3 & 2.2 & 3071.5(6.7) \\
$^{74}Rb$
&  10.42  &  8.3 & 2.6 & 3076(11)\\
\hline
 \end{tabular}
\caption{For 14 superallowed decay channels, the respective $Q$-value, the fractional effect on the decay rate obtained from the energy-dependent correction, the respective shift in the ${\cal F}t$ value, in comparison with the ${\cal F}t$ values and uncertainties taken from \cite{Hardy:2014qxa} are displayed.}
\label{tab1}
\end{table}
It is seen that for the seven most precise ${\cal F}t$ values ($^{26m}$Al through $^{54}$Co) the new correction is comparable with their uncertainties. A systematic, positive sign-definite shift of all ${\cal F}t$ values will then reflect in a substantial shift of their average,
$\overline{{\cal F}t}=3072.27(44){\rm s}\rightarrow\overline{{\cal F}t}=3073.65(46){\rm s}$,
where all uncertainties were treated as statistical, and the uncertainty in $\delta'_R$ was not accounted for (see Ref. \cite{Hardy:2014qxa} for the discussion of its inclusion). However, the new energy-dependent correction is a systematical one, and since I assigned a 100\% uncertainty on its effect on the individual ${\cal F}t$ values, I do the same for the shift of their average, 
\beqn
\left[\delta\overline{{\cal F}t}\right]^{\rm E-dep.}=(1.4\pm1.4)\,{\rm s}.
\eeqn
This effect has always been neglected in the past because it was assumed to be too small. Present analysis shows that that assumption is not justified, and if the relative precision of $2\times10^{-4}$ for the  $\overline{{\cal F}t}$ value and its constancy as a test of CVC and a constraint of non-standard scalar interactions is to be maintained, a robust estimate of this novel effect at the relative 20-30\% or better is necessary. 

Recently, the nuclear modification of the energy-independent correction Re$\Box_{\gamma W}^{even}$ was reevaluated in Ref.~\cite{Seng:2018b}, and this modification was found to be underestimated in the literature \cite{Towner:1994mw,Hardy:2014qxa}. That analysis of the inner correction suggested a shift of a size similar  to that obtained in this work but in the opposite direction, 
\beqn
\left[\delta\overline{{\cal F}t}\right]^{\rm E-indep.}=-(1.8\pm1.2)\,{\rm s}.\label{eq:e-indep}
\eeqn
The two corrections contain the same physics and should be considered jointly. 
When added together, the positive energy-dependent correction cancels the reduction of the energy-independent correction, leaving the central value $\overline{{\cal F}t}$ unchanged but with a larger uncertainty, 
\beqn
\overline{{\cal F}t}=3072(2)\,{\rm s}.
\eeqn 
This cancellation supports the conclusion of Ref. \cite{Seng:2018yzq} that the correct value of $V_{ud}$ extracted from the superallowed nuclear decays is lower than previously thought. While individual shifts are substantial, no firm conclusion on the shift of the central value of $\overline{{\cal F}t}$ with respect to the analysis of Ref. \cite{Hardy:2014qxa} can be made at this stage. The size of the nuclear corrections found in this work and in \cite{Seng:2018b} can be used to estimate the additional uncertainty, and the deficit of the CKM first-row unitarity becomes $|V_{ud}|^2+|V_{us}|^2+|V_{ub}|^2-1=-0.0016(6)$.\\

In summary, I considered a novel effect of a distortion of the electron spectrum in superallowed nuclear $\beta$-decays due to nuclear polarizabilities. This effect has been neglected in the literature based on dimensional arguments originating from the neutron decay. I showed that these arguments are not applicable to nuclear decays where the $Q$-values and nuclear separation energies are of similar size, leading to a slightly higher probability for emitting the electron at the upper end of the spectrum, than at the lower end. 
I estimated the size of the correction to be applied to the ${\cal F}t$ values using a na\"ive dimensional analysis operating with the dipole nuclear polarizability, and in the free Fermi gas model, and demonstrated that the effect is sizable and shifts the resulting average ${\cal F}t$ value towards larger values. On the other hand, the free Fermi gas estimate of the energy-independent nuclear polarizability correction of Ref. \cite{Seng:2018b} led to a shift of the average ${\cal F}t$ value in the opposite direction and of the similar size. Upon incorporating both contributions the $\overline{{\cal F}t}$ value remains roughly unaffected. The exact extent of this cancellation and the size of both effects should be assessed in a more precise way.  Moreover, in this work and in Ref. \cite{Seng:2018b} only the one-nucleon part of the nuclear Green's function was considered. Remaining contributions coming from two-nucleon contributions were taken from Ref. \cite{Hardy:2014qxa} for $\delta_{NS}$, and neglected for $\delta_{NS}^E$. Both contributions are worth an investigation in an upcoming work that should capitalize on recent advances in nuclear theory.

The distortion of the energy spectrum of positrons from nuclear $\beta^+$ decay is a measurable effect. One of the motivations of high-precision measurements of the $\beta$-decay spectra  is the search for new scalar and tensor interactions \cite{Gonzalez-Alonso:2018omy}. The presence of new scalar interactions, {\it e.g.}, would lead to a distortion of the lower part of the spectrum, $\sim b\frac{m_e}{E}$. The energy-dependent effect considered here would enhance the higher part of the spectrum with respect to the analysis of Ref. \cite{Hayen:2017pwg} and beyond its claimed uncertainty, and planned experiments may help confirming or constraining this novel effect. Conversely, experimental searches for Fierz interference in $\beta$ decays may crucially depend on its inclusion: if both distortions go in the same direction, their joint effect on the spectrum may not change its shape, erroneously returning a null result for the Fierz interference if the distortion due to  $\delta_{NS}(E)$ 
is not properly taken into account.

\begin{acknowledgments}
The author acknowledges useful discussions with C.-Y. Seng, M.J. Ramsey-Musolf, J. Hardy, B. Marciano, B. Holstein and Z. Meziani, and 
support by the Deutsche Forschungsgemeinschaft under the Grant No. GO 2604/2-1, and by the German--Mexican collaboration grant SP 778/4--1 (DFG) and 278017 (CONACyT).
\end{acknowledgments}


\begin{thebibliography}{99}

  \bibitem{PDG2018}  M.~Tanabashi et al. (Particle Data Group), Phys.\ Rev.\ D\ {\bf98}, 030001 (2018)

\bibitem{Gonzalez-Alonso:2018omy}
  M.~Gonzalez-Alonso, O.~Naviliat-Cuncic and N.~Severijns,
  Prog.\ Part.\ Nucl.\ Phys.\  {\bf 104} (2019) 165
  doi:10.1016/j.ppnp.2018.08.002
    
\bibitem{Hardy:2014qxa}
  J.~C.~Hardy and I.~S.~Towner,
  Phys.\ Rev.\ C {\bf 91} (2015) no.2,  025501
  doi:10.1103/PhysRevC.91.025501
  
\bibitem{Hardy:2018zsb}
  J.~C.~Hardy and I.~S.~Towner,
  arXiv:1807.01146 [nucl-ex].
  
\bibitem{Sirlin:1967zza}
  A.~Sirlin,
  Phys.\ Rev.\  {\bf 164} (1967) 1767.
  
\bibitem{Marciano:2005ec}
  W.~J.~Marciano and A.~Sirlin,
  Phys.\ Rev.\ Lett.\  {\bf 96} (2006) 032002
  doi:10.1103/PhysRevLett.96.032002
  
\bibitem{Hayen:2017pwg}
  L.~Hayen, N.~Severijns, K.~Bodek, D.~Rozpedzik and X.~Mougeot,
  Rev.\ Mod.\ Phys.\  {\bf 90} (2018) no.1,  015008
  doi:10.1103/RevModPhys.90.015008
  
\bibitem{Seng:2018yzq}
 C.~Y.~Seng, M.~Gorchtein, H.~H.~Patel and M.~J.~Ramsey-Musolf,
  Phys.\ Rev.\ Lett.\  {\bf 121} (2018) no.24,  241804
  doi:10.1103/PhysRevLett.121.241804
  
  
\bibitem{Seng:2018b}   C.~Y.~Seng, M.~Gorchtein and M.~J.~Ramsey-Musolf,
  arXiv:1812.03352 [nucl-th].
  
  
\bibitem{Towner:1994mw}
  I.~S.~Towner,
  Phys.\ Lett.\ B {\bf 333} (1994) 13
  doi:10.1016/0370-2693(94)91000-6
  
\bibitem{DeJager:1974liz}
  C.~W.~De Jager, H.~De Vries and C.~De Vries,
  Atom.\ Data Nucl.\ Data Tabl.\  {\bf 14} (1974) 479
   Erratum: [Atom.\ Data Nucl.\ Data Tabl.\  {\bf 16} (1975) 580].
  doi:10.1016/S0092-640X(74)80002-1

\bibitem{Berman:1975tt}
  B.~L.~Berman and S.~C.~Fultz,
  Rev.\ Mod.\ Phys.\  {\bf 47} (1975) 713.
  doi:10.1103/RevModPhys.47.713
  
  \end{thebibliography}
\end{document}